\documentclass[12pt]{iopart}
\usepackage{iopams}
\usepackage{txfonts}
\usepackage{graphicx}
\begin{document}

\article[Short title]{Star-forming Dwarf Galaxies: Ariadne's Thread in the
  Cosmic Labyrinth}{Models of evolution of gas and abundances in dwarf irregular galaxies}

\author{Gavil\'{a}n, Marta$^1$; Moll\'{a}, Mercedes$^2$ and D\'{i}az, \'{A}ngeles$^1$}

\address{$^1$ Unversidad Aut\'{o}noma de Madrid. Spain}
\address{$^2$ CIEMAT, Madrid, Spain}
\ead{\mailto{marta.gavilan@uam.es}}

\begin{abstract}
We have developed a grid of chemical evolution models applied to dwarf isolated galaxies, using \cite{gav05} yields. The input data enclose different star formation efficiencies, galaxy mass and collapse time values. The result is a wide collection of solutions that vary from objects with low metallicity and great amount of gas, to those with little gas and high metallicity. No environmental effects like tidal or galactic winds have been treated, so these objects are expected to be close to field dwarf galaxies, more than cluster ones. We have studied the time evolution of the abundance of oxygen and nitrogen and the amount of gas, related to their star formation history, as well as the possibility of gas losses by SN winds.
\end{abstract}


\section{Introduction}
According to the hierarchical model, dwarf galaxies are the building blocks of larger structures, consequently their study is crucial to understand galaxy formation and evolution. Although dwarf galaxies are the most numerous objects in the universe, there are some questions that remain apparently unresolved. In fact three of them deal with the same idea about their evolution.
The first question is related to observed mass metallicity relation: Do these objects have a metallicity according to what we can expect?. In other words: Why less massive objects have lower metallicity? Since \cite{leq79} many efforts have been devoted to explain it. The second question is related to a theoretical idea: Are dwarf galaxies able to retain their gas in spite of their shallow potential wells? Although dwarf galaxies are simpler than bigger ones, they exhibit different structures that avoid a simple gravitation calculus. We cannot find the answer only in the scape velocity.
And the third question is related to their star formation history: Are they really young objects? What have they been doing during the last 13 Gyr? 
\section{Models}
The main features of the model are: a) gas outflow is not considered; b) there is a continuous infall of primordial gas from the halo to the galaxy c) the star formation process is continuous, but not constant, so bursts are not taken into account, and d) a galaxy is modeled only as a single region, so no spatial gradients will be obtained.
The model works with a scenario that begins with a primordial gas sphere. The gas begins to fall to the central part, controlled by the collapse time (inverse to the infall rate). In this collapse molecular clouds are formed, and a proportion of them are transformed into stars. These processes are controlled by different efficiencies. The star formation process is continuous, but it is not constant, thus the SFR has different values along time.
Basically the model has two engines: the first one that calculates the chemical evolution, based on the work by \cite{fer92}, in its later version of \cite{mol05}, and the second one that provides the spectroscopic and photometric evolution by \cite{mol08}

All the models use the same stellar yields and SN rates. For massive stars we have chosen those of \cite{woo95} and for low and intermediate mass stars we have selected the ones by \cite{gav05}. The SN rates have been provided personally by Pilar Ruiz-Lapuente.
The things that change from one model to another one are the efficiencies and the collapse time (or the infall rate). To be consistent, we have joined the efficiencies in pairs, in the sense that if there are good conditions to form molecular clouds it can be expected that the conditions would be good for the stars as well. The results are six different values, from low to high efficiency. For the collapse time we have selected four values: 8, 20, 40, and 60 Gyr.
We have run the grid of models for six different galaxy masses, varying from $8 \times 10^{6}M_{\odot}$ to $3 \times 10^{9}M_{\odot}$, representative for the objects under study.
\section{Data}
To be consistent all over this work we will compare our results with a sample of well known galaxies with some prescriptions: a) they have to be dwarf, b) It has to be clearly known if they are isolated or belong to a cluster, c) they have to be gas rich objects (dSph and dE are not considered). With all these prescriptions we have selected a small sample of \textit{good objects}: some dIrr, a few field BCD and some BCD of the Virgo Cluster. dIrr data are from \cite{zee00,zee06a,lee03,gar02, tol00,kar04,ski96}. For the field BCD, we have used data of \cite{cai01,cai03,huc07,nav06,gil03}. For Virgo cluster BCD we have used data of \cite{vil03} and \cite{vad07}.
\section{Results}
\begin{figure}
\centering
\includegraphics[width=10cm]{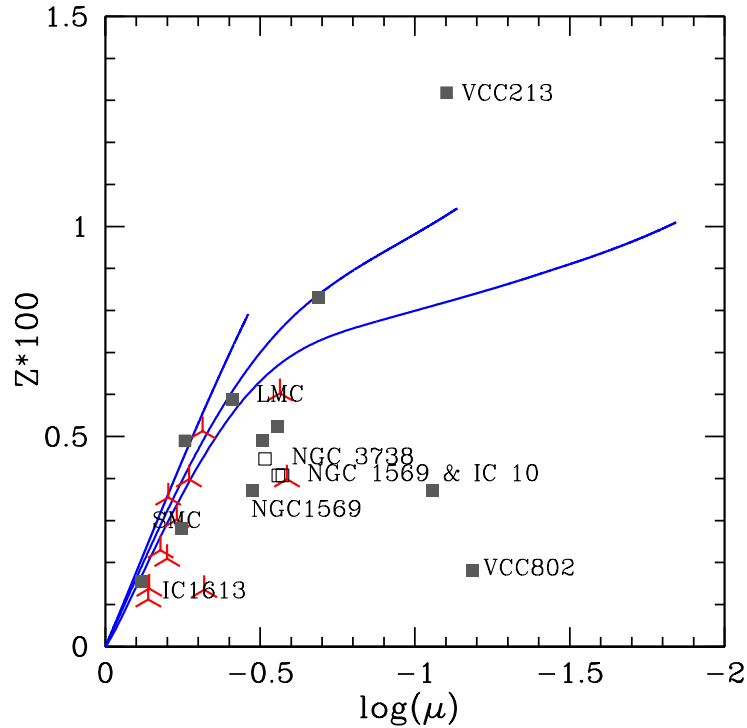}
\caption{Gas fraction \textit{vs} oxygen abundance. dIrr are plotted with stars, isolated BCD with open squares and cluster BCD with solid squares.}
\label{Zmu}
\end{figure}

\paragraph{Gas fraction vs oxygen abundance.}
The gas fraction is the ratio between the mass in gas form and the mass in gas and stars, and in a system that evolve as a closed box, obviously the more mass in stars it has the less the gas fraction will be. But we have to take into account that our model incorporates infall, so the galaxy can increase its stellar mass without an appreciable decrease of the gas. In Fig.~\ref{Zmu} we show the relation between the gas fraction and the metallicity of three models for the most massive galaxy with different star formation efficiencies: extra-low, medium and extra-high.

As can be seen in the figure, dIrr lie nearer the models than BCD, except for NGC 1569, that is considered a \textit{post-star-burst} galaxy, in fact some authors consider it as a dIrr while others classify it as a BCD. 

Cluster BCD are spread in a bigger area than isolated galaxies. Only two objects present extreme values: VCC802 and VCC213. VCC802 is clearly located in the \textit{forbidden} area, the place for objects that have little gas despite their low metallicity. Has this object suffered any interaction during its life? This low metallicity can be due to a gas loss? The case of VCC213 is different, it has too high metallicity although it has a lot of gas. This could be an effect of an enriched infall due to an interaction with other companion, or it could be produced by a violent star burst. In both cases it is difficult to explain these values with models of continuous star formation.

\paragraph{Nitrogen abundance}
In the absence of knowledge of the SFH, nitrogen can give some clues about the age of an object, and is generated in low and mainly intermediate mass stars, and owing to their long life, it returns to the ISM hundreds of mega-years after oxygen does.

\begin{figure}
\centering
\includegraphics[width=10cm]{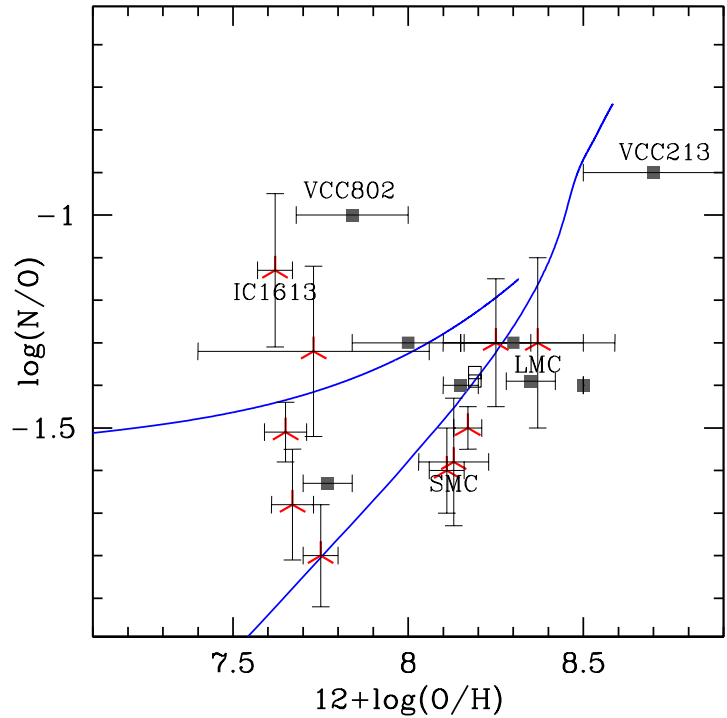}
\caption{(N/O) \textit{vs} (O/H). Only extreme models are represented, the rest of the models lay between them. Symbols are the same than in previous figure. }
\label{NO}
\end{figure}

In Fig.~\ref{NO} the relation between nitrogen and oxygen is plotted. As in the previous case, we only present the extreme cases, and the rest of the models lay between these two curves. Strong bursting star formation histories would produce high oxygen abundances soon; in contrast, a low and continuous SFR keeps the oxygen abundance low for a long time; thus, a large quantity of the primary nitrogen may be ejected.Therefore, the upper line corresponds to the case of a smooth star formation history, thus the primary nitrogen can bee seen in the ISM, and the slope is very flat; the opposite situation is represented by the lower line. In this case, the SF process is closer to a burst, and oxygen increases its abundance very soon, so the slope is similar to a secondary behavior.

Isolated dIrr data are located all over the plot, but taking out IC 1613 most of them are near the model lines. For example, the Magellanic Clouds are very close to the extreme SFR line, corresponding to objects with a young population, or with recent star formation. Other dIrr galaxies are in the transition area. Once again, IC1613 and VCC802 cannot be fitted with the models. Both present very high (N/O) ratio for their poor oxygen abundance. If the galaxy loses gas because of a burst, this gas will be oxygen rich, and when the nitrogen goes to the ISM, it will find less oxygen than can be expected, and consequently N/O will be high.

VCC213 may have different explanation. It has high oxygen abundance, so if it presents high (N/O) is because it has a lot of nitrogen as well. This can be reached through two bursts: the first that has enough time to generate the nitrogen, and the second one, and more recent, that increases the oxygen, but has not had time to expel the nitrogen.

\begin{figure}
\centering
\includegraphics[width=10cm]{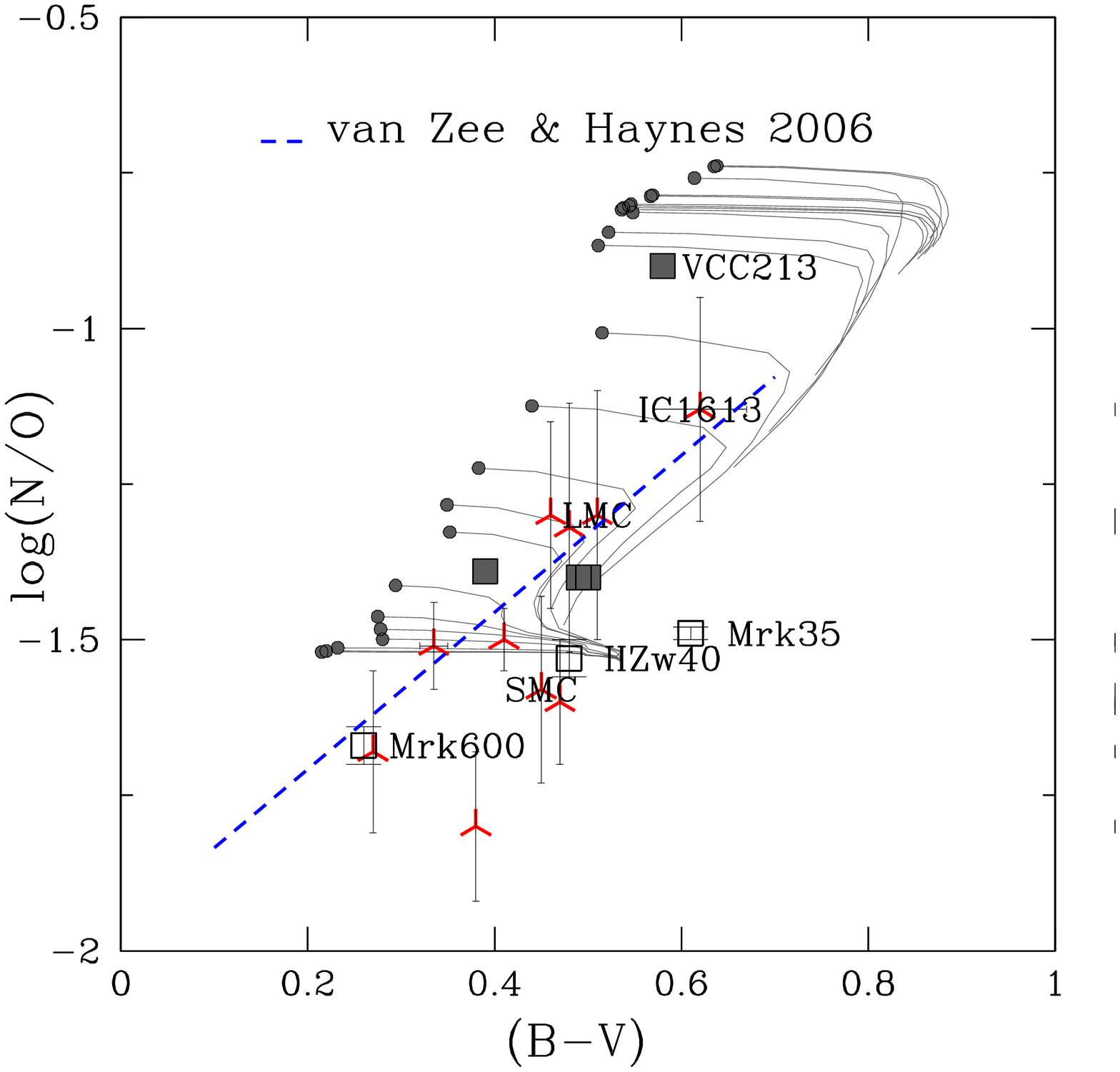}
\caption{Evolution tracks of $M_{B}$ vs (N/O) relation. Dashed line corresponds to the relation obtain by \cite{zee06}. Grey dots are the final points of each model.}
\label{COL_NO}
\end{figure}

\paragraph{Nitrogen vs color}
Following the idea of \cite{zee06} about the relation between (N/O) and color, in Fig~\ref{COL_NO} we present this relation for our models and the data. Solid circles are the final values reached by models, and they would match the data if the age of the objects would be 13.2 Gyr, in other cases, the data will be located between the lines, as the majority of dIrr do. This is not the case of isolated BCD galaxies, because they have lower (N/O) abundance, so they are located in the lower part of the plot. Probably, models with star formation bursts would be needed to match them.

Cluster BCD have a different behavior from that of isolated ones and, as can be seen in the plot, they fit the models better. Nevertheless VCC213 is bluer than the general tendency. This can be due to a second burst population. This population has to be very young, since it has not had time to eject its nitrogen, so its color should be blue.

\begin{figure}
\centering
\includegraphics[width=10cm]{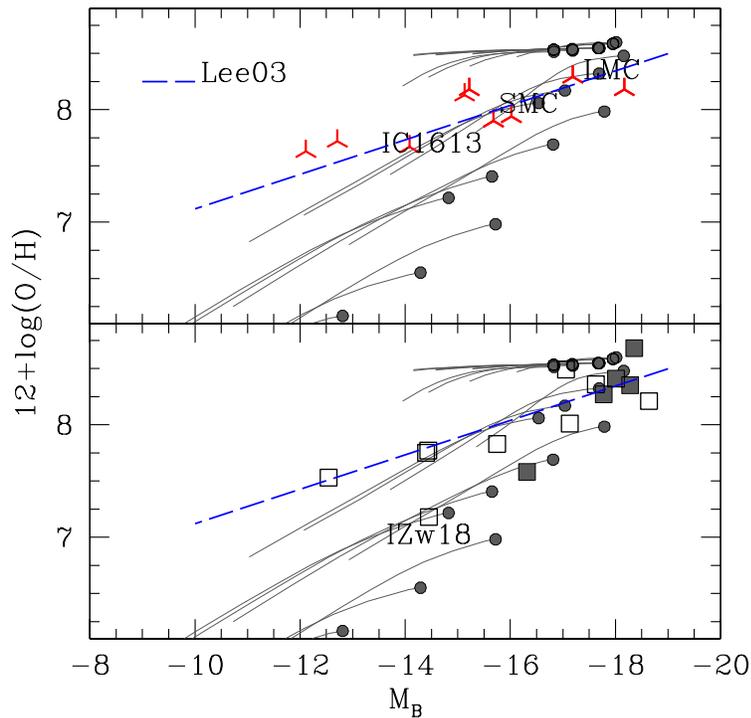}
\caption{Luminosity-metallicity relation evolution tracks. Dashed line is the relation calculated by \cite{lee03} for nearby dwarf galaxies. Symbols are the same than in previous figures.}
\label{LbOH}
\end{figure}

\paragraph{Luminosity-metallicity relation}
Finally, in Fig~\ref{LbOH} we show the luminosity-metallicity relation, as a way to estimate the \cite{leq79} relation. In order to have a clearer plot, we have separated dIrr data (in the upper panel) from BCD (in the lower panel). Lines show time evolution of the relation, and solid circles correspond to the final values. The dashed line shows the relation found by \cite{lee03} for nearby dwarf galaxies with [OIII] lines detected.

In the case of dIrr, the slope shown by the data is flatter than what we obtain with the models, wich in fact, do not reproduce the relation calculated by \cite{lee03}. The models with higher efficiencies present a very flat evolution track. They reach high oxygen abundances very soon and, as they decrease their SFR along time, the (O/H) level remains constant, but this is not a realistic scenario for dwarf irregular galaxies. We find a better fit with models with lower efficiencies, and in these cases, the tracks followed by them are similar to the empirical relation.

In BCD we find again different trends in isolated and cluster BCD. Isolated BCD show a slope very similar to dIrr, that is flatter than our models, but cluster BCD agree quite well with the final points of the evolution. We have to consider that only five objects are represented. More cluster BCD data would be needed to obtain robust conclusions.

\section{Conclusions}

From our results of gas, metallicity and photometry, we can conclude that most isolated dIrr galaxies can be modeled by continuous star formation processes, if primordial gas infall is considered. In this scenario most of them seem to be young objects.

This is not the case for BCD, neither isolated nor cluster ones, but it is surprising that more coincidences between models and data have been found for cluster BCD than for isolated ones, mainly for color results.

\end{document}